\input epsf.tex
\documentstyle[12pt]{article} \textwidth=6in \oddsidemargin=0in
\begin{document}
\newcommand\bfigt{\begin{figure}[top]}
\newcommand\efig{\end{figure}}
\newcommand{\dl}{\delta}
\newcommand{\iy}{\infty}
 \newcommand{\La}{\Lambda}
\newcommand{\pl}{\partial}
 \newcommand{\noi}{\noindent}
\renewcommand{\sp}{\vskip2ex}
\newcommand{\bq}{\begin{equation}}
\newcommand{\eq}{\end{equation}}
\newcommand{\ra}{\rightarrow}
\newcommand{\al}{\alpha}
\newcommand{\th}{\theta}
\newcommand{\bR}{{\bf R}}
\newcommand{\bC}{{\bf C}}
 \newcommand{\bZ}{{\bf Z}}
\newcommand{\nm}{\parallel}
 \newcommand{\la}{\lambda}
  \newcommand{\ph}{\varphi}
\newcommand{\om}{\omega}
\newcommand{\inv}{^{-1}}
\newcommand{\ve}{\varepsilon}
\newcommand{\Ga}{\Gamma}
\newcommand{\hf}{{1\over 2}}
\newcommand{\cH}{\cal H}
\newcommand{\ch}{\raisebox{.4ex}{$\chi$}}
\newcommand{\ov}{\over}
\newcommand{\Om}{\Omega}
\newcommand{\ba}{\left(\begin{array}{cc}}
\newcommand{\ea}{\end{array}\right)}
\newcommand{\cM}{{\cal M}}
\newcommand{\be}{\beta}
\newcommand{\uv}{\Big({u\ov v}\Big)^{\be}}
\renewcommand{\sl}{s_{\la}}
\newcommand{\ub}{\,u^{\be}}
\newcommand{\um}{\,u^{-\be}}
\newcommand{\tl}{\widetilde}
\newcommand{\ds}{\displaystyle}
\newcommand{\ssl}{\textstyle}
\newcommand{\ga}{\gamma}

\begin{center}
{ \large\bf  On Exact Solutions to the Cylindrical Poisson-Boltzmann }
\end{center}
\begin{center}{\large\bf Equation with Applications to
Polyelectrolytes\footnotemark[1]}
\footnotetext[1]{Dedicated to Benjamin Widom on the occasion of his seventieth
birthday.}
\end{center}
\sp\begin{center}{{\bf Craig A. Tracy}\\
{\it Department of Mathematics and Institute of Theoretical Dynamics\\
University of California, Davis, CA 95616, USA\\
e-mail address: tracy@itd.ucdavis.edu}}\end{center}
\begin{center}{{\bf Harold Widom}\\
{\it Department of Mathematics\\
University of California, Santa Cruz, CA 95064, USA\\
e-mail address: widom@math.ucsc.edu}}\end{center}\sp
\begin{abstract}
Using exact results from the theory of completely integrable systems of
the Painlev{\'e}/Toda type, we examine the consequences for the theory of
polyelectrolytes in the (nonlinear) Poisson-Boltzmann approximation.
\end{abstract}
\renewcommand{\theequation}{1.\arabic{equation}}
\noi{\bf 1. Introduction}\sp
A polyelectrolyte is a macromolecule with a large number of ionizable groups
which
 in a solvent   becomes highly charged~\cite{oosawa,barrat}.
This highly charged
structure is referred to as the  polyion.
If the polyion has charge $Q=Z e$, then in solution
there are $Z$ counterions.  (We take the polyion to have negative charge
 and thus the counterions
have positive charge.) Typically the solvent also contains a salt and so the
total number
of counterions is $Z$ plus the number of positive ions from the salt.  The
negative salt
ions are called coions. Typically the excess salt is a 1--1 salt, e.g.\ NaCl,
though other
salts are considered.  ($\mbox{MgCl}_2$ is an example of a 2--1 salt.)
 \par
  One class of polyelectrolytes
 that is widely studied are those consisting of slender, rod-like particles.
For example,
 the tobacco
 mosaic virus, a polyelectrolyte, has a diameter of about 18 nm with a length
of 300 nm.  For such
 systems the idealized model is an infinite cylinder of radius $a$ of
  uniform linear charge density with counterions and coions treated as point
  particles.  This neglects several important effects such as the interaction
  between polyions, the flexibility degrees of freedom of the polyion, finite
size effects, to name
 but a few.  Nevertheless, this model has been
 extensively studied (for reviews see~\cite{anderson1,anderson2}).
  In this idealized model there are two additional
 length scales: the electrostatic
 length scale (Bjerrum length) $\ell_B=e^2/\ve k_B T$  ($\ve$ is the solvent
permittivity)
 and the inverse Debye-H{\"u}ckel screening parameter $\kappa^2=8\pi\ell_B I$
 where $I={1\over 2}\sum n_j q_j^2$ is the ionic strength
 with  $n_j$  the molar concentration of ion $j$ of (integer-valued) charge
$q_j$ with
 the convention that $q_j<0$ for counterions and $q_j>0$ for coions.
 From these two lengths we
 form the dimensionless {\it Manning parameter\/}
 \cite{manning1,manning2}
 $ \xi=\ell_B/ b $
 where $b$ is the average spacing between charges on the polyion and the
 dimensionless distance parameter
 $ r=\kappa R $ where $R$ is the cylindrical distance variable.
 \par
 A  mean field theory approach to this model results
 in  the Poisson-Boltzmann (PB)
 equation; that is, the electrostatic potential $\Psi$ is assumed to satisfy
 the Poisson equation of electrostatics with the density of various ions being
given in terms of
  Boltzmann factors \cite{anderson1,anderson2,katchalsky,fixman, qian}.
  Introducing the reduced
   potential $y=e\Psi/k_B T$ the PB equation
 becomes
 \[ \Delta y = - 4\pi \ell_B \sum_j q_j n_j e^{-q_j y}\, . \]
 In terms of  $r$ and for cylindrical geometry
 the PB equation  is, explicitly for the cases of  1--1  and 2--1
  ``excess of salt'' \cite{katchalsky},
 \bq
 {d^2 y\ov d r^2} + {1\ov r} {d y\ov d r} = \left\{ \begin{array}{ll}
 					\sinh y & \mbox{1--1 salt,} \\
 					{1\ov 3}\left(e^{2y}-e^{-y}\right) & \mbox{2--1 salt.}
 					\end{array} \right. \label{PB}\eq
 \par
One boundary condition for  (\ref{PB}) is obtained  by applying Gauss's Law
at the surface of the polyion
\bq  \lim_{r\ra a}\ r {dy\ov dr}  = -2\xi . \label{BC1} \eq
(We use the same symbol $a$ for both the polyion radius and the dimensionless
polyion radius.)\   The second boundary condition depends upon whether the
system is
closed or open.  For a closed system one requires that the electric field
vanish at some
finite distance whereas for an open system with a finite concentration of added
salt one
requires that
\bq y(r) \ra 0 \mbox{ as } r \ra \iy . \label{BC2} \eq
 We will consider only the case of an open system.
A further simplification, and one we will also assume, is to take the polyion
radius $a\ra 0$
which means we model the polyion as a line charge.
The mathematical problems that we address in
this paper are now well formulated: To solve (\ref{PB})  subject to the
boundary
conditions (\ref{BC1}) and (\ref{BC2}).
\sp
\setcounter{equation}{0}\renewcommand{\theequation}{2.\arabic{equation}}
\noi{\bf 2. Exact Solutions  for a 1--1 and 2--1  Salt}\sp
First note that the solutions to the linearized versions of (\ref{PB})
satisfying
(\ref{BC1}) and (\ref{BC2}) are in both cases the familiar Debye-H{\"u}ckel
solution
\bq y_{DH}(r) = 2\xi K_0(r) \label{DH} \eq
where $K_0$ is the modified Bessel function.
The exact solution for the 1--1 salt  is \cite{mtw,tw7,w1}
\begin{eqnarray}
y_{11}(r) & = & 2 \log\det\left(I+\la K\right) - 2 \log\det\left(I-\la
K\right)\nonumber\\
& = &4 \sum_{j=0}^\iy {\la^{2j+1}\ov 2j+1} \mbox{Tr}\left(K^{2j+1}\right)
\label{y11}
\end{eqnarray}
where $K$ is the integral operator on ${\bf R}^+$ with kernel
\[ {\exp\left(-r/2(x+1/x)\right)\ov x+y } \]
and
\[ \la = {1\ov \pi} \sin{\pi\xi\ov 2} \mbox{ \quad for } \xi\le 1. \]
Note that $\mbox{Tr}(K)=K_0(r)$ and that $\mbox{Tr}(K^j)=\mbox{O}(e^{-jr})$
as $r\ra\iy$.
For  $r\ra 0$ it has been proved that \cite{mtw,tw9}
\bq
y_{11}(r) = -2\xi\log r +6\xi\log 2
+ 2 \log{\Ga\left({1+\xi\ov 2}\right)\ov\Ga\left({1-\xi\ov 2}\right)}
 + \mbox{o}(1) \mbox{ \quad for } \xi<1 \label{smallr11}
\eq
where $\Ga$ is the gamma function.
\par
The exact solution for the 2--1 salt  is \cite{w1}
\begin{eqnarray}
y_{21}(r) & = &  \log\det\left(I-\la K_3\right) -  \log\det\left(I-\la
K_2\right)\nonumber\\
	& = & \sum_{j=0}^\iy {\la^{j}\ov j}\left(\mbox{Tr}\left(K_2^j\right)-
	\mbox{Tr}\left(K_3^j\right)\right)
\label{y21}\end{eqnarray}
where  $K_j$ ($j=2,3$)  are integral operators on ${\bf R}^+$ with kernel
\begin{eqnarray*}
&\om^{j-2} (1-\om)
\exp\left[-(r/2\sqrt{3})\left((1-\om)x+(1-\om^{-1})x^{-1}\right)\right]\,
	\left(-\om x + y\right)^{-1}  \\
&	+ \>\om^{2(j-2)}(1-\om^2)
	\exp\left[-(r/2\sqrt{3})\left((1-\om^2)x+(1-\om^{-2})x^{-1}\right)\right]\,
	\left(-\om^2 x + y\right)^{-1}.
\end{eqnarray*}
Here $\om=e^{2\pi i/3}$ and
\[ {\la\ov \la_c} = 2\sin\left({2\pi\ov 3}(\xi+1/4)\right) -1,\> \la_c={1\ov 2
\sqrt{3}\pi},
\mbox{ \quad for } \xi\le {1\ov 2}. \]
Again note that
\bq \mbox{Tr}\left(K_2\right)-\mbox{Tr}\left(K_3\right) = 6
K_0(r).\label{Tr1}\eq
An elementary calculation produces a  simpler representation than the
definition for the trace of the square
of the operators:
\bq \mbox{Tr}\left(K_2^2\right)-\mbox{Tr}\left(K_3^2\right) =9\int_0^\iy
\int_0^\iy
{\exp\left(-r/2(x_1+1/x_1+x_2+1/x_2)\right)\ov x_1^2+x_1 x_2 + x_2^2 }\, dx_1
dx_2\, .
\label{Tr2}\eq
The computations (\ref{Tr1}) and (\ref{Tr2})  lead one to suspect that
 for all postive integers
$n$ the quantities $\mbox{Tr}(K_2^n)-\mbox{Tr}(K_3^n)$, as functions
of $r$, are monotonically decreasing.
 Indeed, one can derive an alternative matrix kernel
representation of the operators $K_j$ such that
  the monotonic decay  is manifestly
clear.
\par
For  $r\ra 0$ it has been proved that \cite{tw9,kitaev}
\bq
y_{21}(r) = -2\xi\log r + \left(2\log 2 + 3\log 3\right)\xi +\log{
\Ga\left({1+\xi\ov 3}\right)\Ga\left({2+2\xi\ov 3}\right)\ov
\Ga\left({2-\xi\ov 3}\right)\Ga\left({1-2\xi\ov 3}\right)}
+ \mbox{o}(1) \mbox{ \quad for } \xi<1/2. \label{smallr21}
\eq
Higher order terms in both (\ref{smallr11}) and (\ref{smallr21}) can be
computed
from use of the differential equations and all additional constants appearing
can
be expressed in terms of the quantities above.
\sp
\setcounter{equation}{0}\renewcommand{\theequation}{3.\arabic{equation}}

\noi{\bf 3. Asymptotics at the Critical Value of $\xi$---Counterion
Condensation}\sp
The Oosawa-Manning arguments \cite{oosawa,manning1,manning2}
 for counterion condensation are well known and need
not be repeated here.  These arguments, when applied to a 1--1 salt and 2--1
salt, predict critical values of $\xi$ at $1$ and $1/2$, respectively.  Recall
that
in the theory of counterion condensation, the meaning of $\xi>\xi_c$ is that
the average
charge spacing $b$ on the polyion
 is increased by counterion condensation (onto or around the polyion)
until $\xi=\xi_c$ is achieved.  Thus one must distinguish between the
stoichiometric
value of $\xi$ computed using only the charge groups on the polyion and the
lower
value of $\xi$ achieved through counterion condensation.
\par
Mathematically, the critical value of $\xi$ is seen through the qualitative
change in the small $r$ asymptotics (\ref{smallr11}) and (\ref{smallr21}).  For
the 1--1
salt case we have as $r\ra 0$ \cite{mtw,tw9}
\bq
\exp\left(-y_{11}(r)/2\right)=-{r\ov 2}\Om_1 - {r^5\ov
2^{12}}\left(8\Om_1^3-8\Om_1^2
+4\Om_1-1\right) + \mbox{O}(r^9\Om_1^5) \label{rsmall11C}\eq
where
\[\Om_1=\Om_1(r)=\log(r/8)+\ga\]
and $\ga$ is Euler's constant.
Thus the potential $y_{11}(r)$ develops an additional $\log\log r$ singularity
at
the critical Manning parameter.
Similarly for the 2--1 case as $r\ra 0$ \cite{tw9,kitaev}
\bq
\exp\left(-y_{21}(r)\right)=-{r\ov\sqrt{3}}\Om_2+{r^4\ov 81}\left(\Om_2^2-{2\ov
3}
\Om_2+{2\ov 9}\right) + \mbox{O}(r^7\Om_2^3)\label{rsmall21C}\eq
where
\[\Om_2=\Om_2(r)=\log(r/8)+\ga+{1\ov 3}\log 2 + {3\ov 2}\log 3.\]
\par
For $\xi>\xi_c$ the above solutions are no longer physically valid.  For
example,
the Boltzmann factor $\exp\left(-y_{11}(r)\right)$ will become negative for
small enough $r$.
\sp
\setcounter{equation}{0}\renewcommand{\theequation}{4.\arabic{equation}}

\noi{\bf 4. Exact Electrostatic Free Energy for 1--1 and 2--1  Salt Cases}\sp
At constant temperature and pressure the free energy is the work done in
placing
charges on the polyion.  This is the familiar ``charging process'' and one
imagines
an increment of charge $dq$ placed at the surface of the polyion so that the
infinitesimal work done is $\Psi(a) dq$ where the electrostatic potential is
evaluated at $R=a$ \cite{hill,tanford}. Thus
\[w^{\mbox{el}}=\int_0^Q \Psi(a)\, dq \]
is the free energy associated with a single line charge.
Our solutions $y_{11}$ and $y_{21}$ are  of the form
\[ y(r) = -2\xi\log r +y_0(\xi) + \mbox{o}(1) \mbox{ \quad as } r\ra 0\]
and are for the limiting case of $a\ra 0$.  We take, therefore, for the value
of the $e\Psi(a)/k_B T$ the quantity $-2\xi\log(\kappa a) +y_0(\xi)$.
Re-writing the
above expression for $w^{\mbox{el}}$ in terms of dimensionless quantities and
multiplying
the result by $N_p$, the total number of polyions in solution of volume $V$, we
obtain for the free energy, $W^{\mbox{el}}$, of the entire solution
\cite{manning1,manning2}
\begin{eqnarray*}
 {W^{\mbox{el}}\ov V k_B T}&=& -n_p\left(\log(\kappa a) \xi -{1\ov
\xi}\int_0^\xi y_0(\xi')\,
d\xi' \right) \\
&:=& - n_p f(\xi)
\end{eqnarray*}
where $n_p=N_p/V$ is the polyion concentration.
Using the expressions (\ref{smallr11}) and (\ref{smallr21}) we calculate
$f(\xi)$ in
these two cases:
\begin{eqnarray}
f_{11}(\xi)&=&\left(\log(\kappa a)-3\log 2\right)\xi +{2\ov \xi}
\int_0^\xi \log{\Ga(1/2-\xi'/ 2)\ov \Ga(1/2+\xi'/ 2)}\, d\xi' \nonumber \\
&=&\left(\log(\kappa a)-\log 2 +\ga\right)\xi-\sum_{n=1}^\iy
{\psi^{(2n)}(1/2)\ov
2^{2n-1} (2n+2)!} \xi^{2n+1} \nonumber \\
\noalign{\vskip 5pt}
&=&\left(\log(\kappa a) +1 -3 \log 2\right) +2 \log{\Ga({1-\xi\ov 2})\ov
\Ga({1+\xi\ov 2})}
\nonumber \\
\noalign{\vskip 2pt}
&& + {2\ov\xi}\log{\Ga\left({1-\xi\ov 2}\right)\Ga\left({1+\xi\ov 2}\right)\ov
 \Ga^2\left(1/2\right)}
+{4\ov\xi} \log{G({1-\xi\ov 2})G({1+\xi\ov 2})\ov G^2(1/2)},\\
\noalign{\vskip 6pt}
f_{21}(\xi)&=&\left(\log(\kappa a)-\log 2 -{3\ov 2}\log 3\right)\xi +{1\ov\xi}
\int_0^\xi \log{ \Ga\left(2/3-\xi'/3\right)\Ga\left(1/3-2\xi'/3\right)\ov
\Ga\left(1/3+\xi'/3\right) \Ga\left(2/3+2\xi'/3\right)}
\, d\xi' \nonumber\\
\noalign{\vskip 5pt}
&=&\left(\log(\kappa a)-\log 2 +\ga\right)\xi+{1\ov
18}\left(\psi^{(1)}(1/3)-\psi^{(1)}(2/3)\right)
\xi^2\nonumber\\
&& -{1\ov 72}\left(\psi^{(2)}(1/3)+\psi^{(2)}(2/3)\right)\xi^3
+\mbox{O}(\xi^4).
\end{eqnarray}
\bfigt\vspace*{-35mm}\hspace*{25mm}\epsfysize=140mm \epsffile{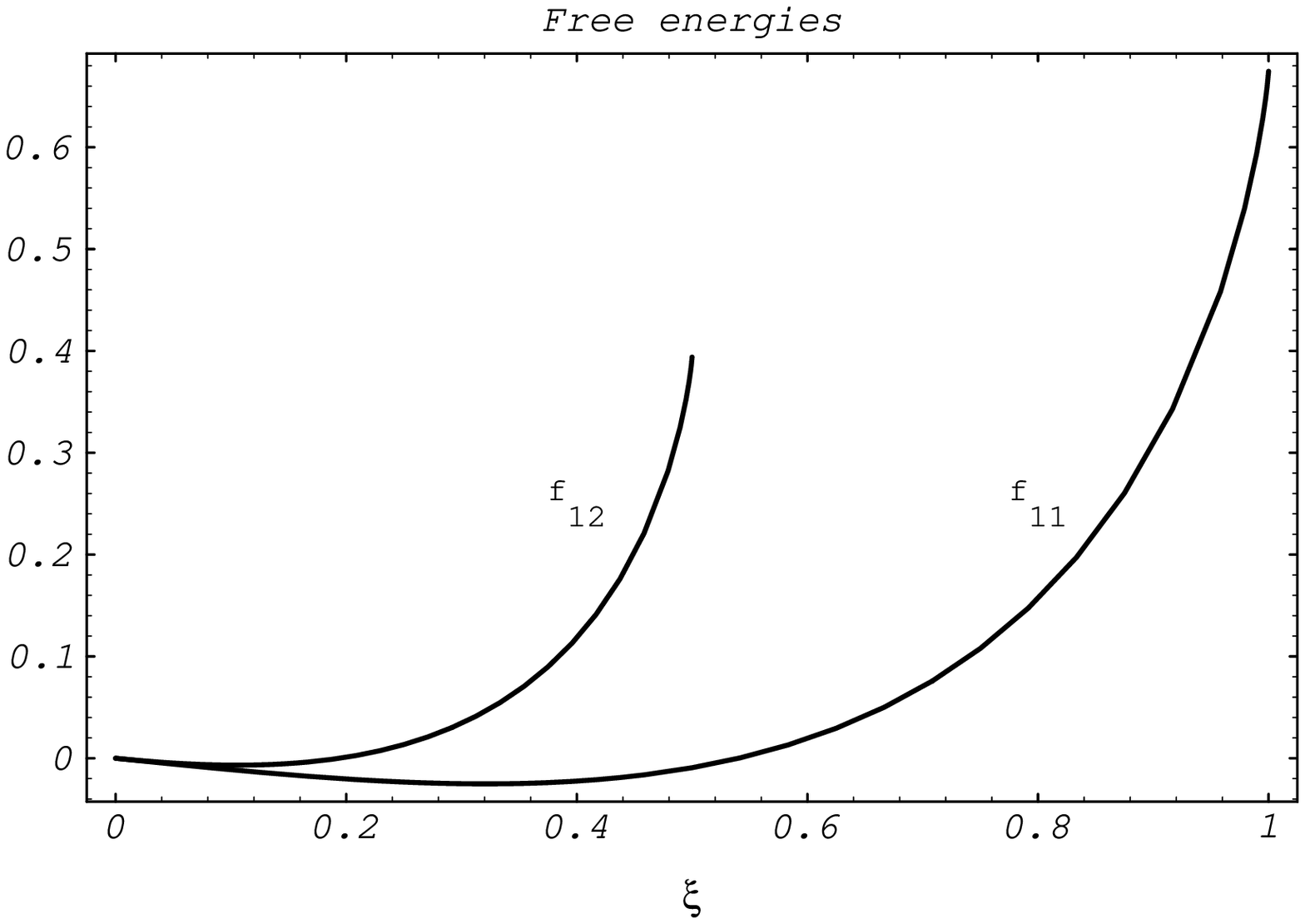}
\vspace*{-35mm}
\caption{The free energies $f_{11}$ and $f_{21}$ as functions of the
Manning parameter $\xi\le \xi_c$
with $\kappa a =1$. At the critical value,  $f_{11}(1)=\log\kappa a +0.674563$
and $f_{21}(1/2)={1\ov 2}\log\kappa a + 0.393915$.}
\efig
In the above $\psi(x) = \Ga^\prime(x)/\Ga(x)$ and $\psi^{(n)}(x)$ is the
$n^{\mbox{th}}$
derivative of $\psi$.  The function $G(s)$ is the Barnes $G$-function
\cite{barnes} which is an entire
function of $s$ and is defined by
\[ G(s+1)=(2\pi)^{s/2} \exp\left(-s/2-(1+\ga)s^2/2\right)\prod_{k=1}^\iy
\left(1+{s\ov k}\right)^k \exp\left(-s+{s^2\ov 2k}\right). \]
The $G$-function satisfies the functional equation $G(s+1)=\Ga(s) G(s)$ and has
the special value $G(1)=1$.  It arises in the present context through the
integral
\[\int_0^z\log\Ga(x+1)\,dx = {z\ov 2}\log 2\pi
-{1\ov 2}z(z+1)+z\log\Ga(z+1)-\log G(z+1).\]
(Clearly one can also express $f_{21}$ in terms of the $G$-function.)
\setcounter{equation}{0}\renewcommand{\theequation}{5.\arabic{equation}}
\par
In the small $\xi$ expansion for both $f_{11}$
and $f_{21}$ the linear term is the   contribution from
the  Debye-H{\"u}ckel theory.  Observe that since the
volume dependence resides solely in the
$\log\kappa a$ term, such derived quantities as the osmotic pressure (which is
expressible
in terms of a first
derivative of $W^{\mbox{el}}$ with respect to $V$) will be identical to that
derived
from the Debye-H{\"u}ckel theory.  The mean activity coefficients $\ga_{\pm}$,
expressible
in terms of the function $f$, will
have different numerical values from the Debye-H{\"u}ckel theory, however their
dependence on $n_p$ will be identical.  Though this has been understood
and used by previous workers, we stress these facts since it emphasizes the
wider
validity
of the linear theory than one might {\it a priori\/}  expect.
\par
Both $f_{11}$ and $f_{21}$ are singular at the critical Manning parameter and
have leading singularities of the form $ (\xi_c-\xi)\log(\xi_c-\xi)$.  The
graphs
of $f_{11}$ and $f_{21}$ are shown in Fig.~1 where
for convenience we have set $\kappa a=1$.
\sp
\noi{\bf 5. Partial Equilibrium Structure Factors }\sp
Within the Poisson-Boltzmann approximation the  density distributions
 of counterions ($+$) and coions ($-$) are given by
 \[ n_{\pm}(r)=n_\pm \exp\left(\mp q_j y(r)\right)\]
  where $n_\pm$ are
 the concentrations at infinity.   Light scatters
 from the local concentration fluctuations of the counterions and coions and
the
 observed scattering intensity in the static approximation
  is expressible in terms of the partial structure
 factors which are, in the cylindrical PB approximation, the (two-dimensional)
Fourier transform
 of $\left(n_{\pm}(r)/n_\pm - 1\right)$ \cite{hermans,berne,odijk}.
 Thus we examine, for  the 1--1 salt,
 \bq
 S_\pm(q,\xi) = {\cal F}(e^{\pm y}-1)=2\pi \int_0^\iy J_0(qr) \left(e^{\pm
y(r)}-1\right) r\,dr
 \label{S}\eq
 where $q$ is the dimensionless wave number and $J_0$ is the Bessel
function
 of zeroth order.
 For the 2--1 salt  we replace $e^y\ra e^{2y}$ in the above expression.
 \par
 At $q=0$,
 \begin{eqnarray}
 S_+(0,\xi)&=&\left\{ \begin{array}{ll}
 		\; 4\pi\xi(1+\xi/2) & \mbox{1--1 salt,} \\
 		\; 8\pi\xi(1+\xi/2) & \mbox{2--1 salt,}
 		\end{array}\right. \label{S0+} \\
 \noalign{\vskip 6pt}
 S_-(0,\xi)&=&\left\{ \begin{array}{ll}
 		\! -4\pi\xi(1-\xi/2) & \mbox{1--1 salt,} \\
 		\! -4\pi\xi(1-\xi) & \mbox{2--1 salt.}
 		\end{array}\label{S0-}\right.
 \end{eqnarray}
 The differences $S_+(0,\xi)-S_-(0,\xi)$ follow directly from the differential
equations
 (\ref{PB}) and the boundary condition (\ref{BC1}).  The individual terms
require
 additional identities (for 1--1 case see eq.~(6) in \cite{tw7}).  In the
Debye-H{\"u}ckel
 approximation (\ref{DH}), $S_\pm(0,\xi)$ is $\pm 4\xi$ for the 1--1 salt and
similarly
 for the 2--1 salt. For the 1--1 salt, $S_-(q,\xi)=S_+(q,-\xi)$.
 \par
 To understand further how these results differ from the Debye-H{\"u}ckel
 approximation, we expand the exponentials in (\ref{S}) and use the expansions
(\ref{y11}) and (\ref{y21})
 to deduce the expansions:
 \bq
 S_\pm(q,\xi)=\sum_{j=1}^\iy \la^j S_{\pm,j}(q). \label{S_exp}\eq
For the 1--1 salt  we
find, for example,
\begin{eqnarray*}
S_{\pm,1}(q) &=&\pm{\cal F}\left(4\mbox{Tr}(K)\right)=\pm{8\pi \ov 1 + q^2}, \\
S_{\pm,2}(q)&=&{\cal F}\left(8\mbox{Tr}(K)^2\right)=
 8\pi {\log\left(q/2+\sqrt{(q/2)^2+1}\right)\ov (q/2) \sqrt{(q/2)^2+1} }\\
 S_{\pm,3}(q)&=&\pm {\cal F}\left( {4\ov 3}(
\mbox{Tr}(K^3)+8\mbox{Tr}(K)^3)\right),\\
 \noalign{\vskip 4pt}
 &=&\pm{32\pi\ov 3}\int_0^\iy \int_0^\iy \int_0^\iy
{(x_1+x_2+x_3)(1/x_1+1/x_2+1/x_3)
 \ov (x_1+x_2)(x_2+x_3)(x_3+x_1)}\\
 \noalign{\vskip 3pt}
 && \quad \times
 {(x_1+1/x_1+x_2+1/x_2+x_3+1/x_3)\ov
  \left[4q^2+(x_1+1/x_1+x_2+1/x_2+x_3+1/x_3)\right]^{3/2} }\, dx_1 dx_2 dx_3.
\end{eqnarray*}
\bfigt\vspace*{-100mm}\hspace*{-18mm}\epsfysize=250mm \epsffile{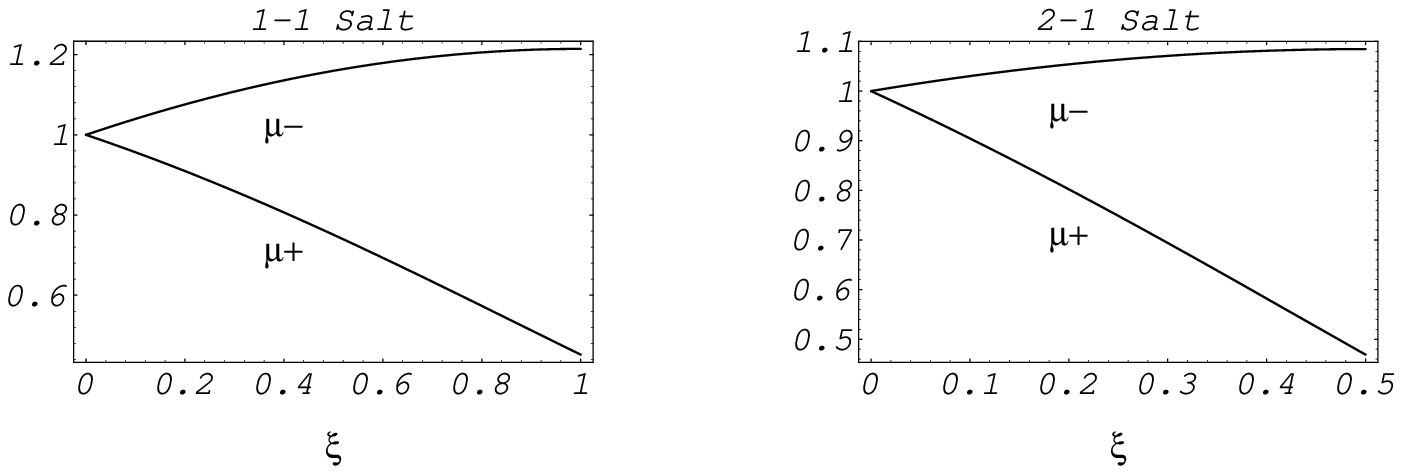}
\vspace*{-100mm}
\caption{The normalized  moments  $\mu_\pm$ as functions of the Manning
parameter
$\xi$.}
\efig
{}From (\ref{S0+}) it follows that $S_{\pm,3}(0)=\pm 4\pi^3/3$.
In general $S_{\pm,2p+1}$ and $S_{\pm,2p+2}$ will involve
the  $\mbox{Tr}(K^{2\ell+1})$
($\ell=1,\ldots,p$) and they will have, in the complex $q$-plane, branch point
singularities at
$q=\pm i p$. Since $\la=\mbox{O}(\xi)$, $\xi\ra 0$,  we see,
for  $\xi\ll 1$, that the Debye-H{\"u}ckel approximation of taking the first
term
in the sum (\ref{S_exp}) is valid for bounded
$q$.
Similar expansions can be derived for the 2--1 salt where we note that
$S_{\pm,1}(q)$
will again be the Debye-H\"uckel term (see (\ref{Tr1})) but now $S_{\pm,2}(q)$
will contain
irreducible ``two-particle'' effects (see (\ref{Tr2})).
\begin{figure}\vspace*{-30mm}\hspace{25mm}\epsfysize=130mm \epsffile{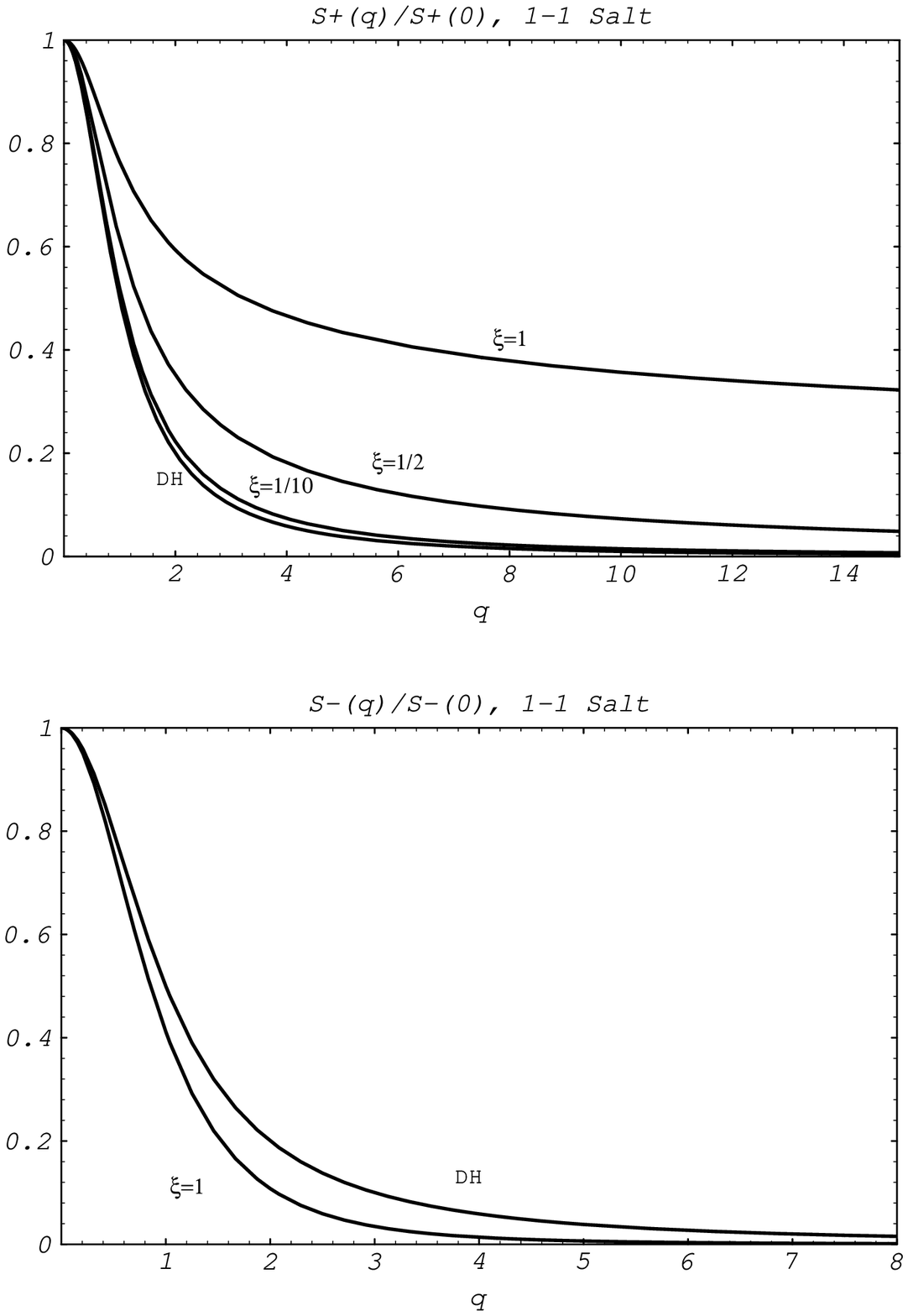}
\vspace*{-10mm}
\caption{The structure factors $S_\pm(q,\xi)/S_\pm(0,\xi)$ as functions of $q$
for
the 1-1 salt.
Also plotted is the Debye-H{\"u}ckel approximation.}
\efig

\par
For any scattering intensity $I(q)$, one   experimentally accessible quantity
is
the normalized
second moment
$\mu$ in the small $q$ expansion
\[ \left[{I(q)\ov I(0)}\right]^{-1}=1+\mu q^2 + \mbox{O}(q^4). \]
For the partial structure factors $S_\pm$ one similarly
defines  $\mu_\pm(\xi)$ and in Fig.~2 we graph $\mu_\pm(\xi)$.
  We have defined the  inverse correlation length by the distance to
  the nearest pole in $S(q)$.
 (The choice of the dimensionless wave number $q=k/\kappa$ fixes
 the location to $\pm i$.)  For the Debye-H{\" u}ckel approximation
 $\mu_\pm\equiv 1$, in contrast with the  PB equation where
  $\mu_+$ ($\mu_-$) decreases (increases) with increasing $\xi$.
 The partial structure
functions themselves are shown in Figs.~3 and 4  for various values of $\xi$.
Note
that with
increasing $\xi$ there is a significant increase in $S_+(q,\xi)/S_+(0,\xi)$
for large $q$ whereas $S_-(q,\xi)/S_-(0,\xi)$ shows only a slow decrease for
increasing $\xi$.
\begin{figure}\vspace*{-30mm}\hspace{25mm}\epsfysize=130mm \epsffile{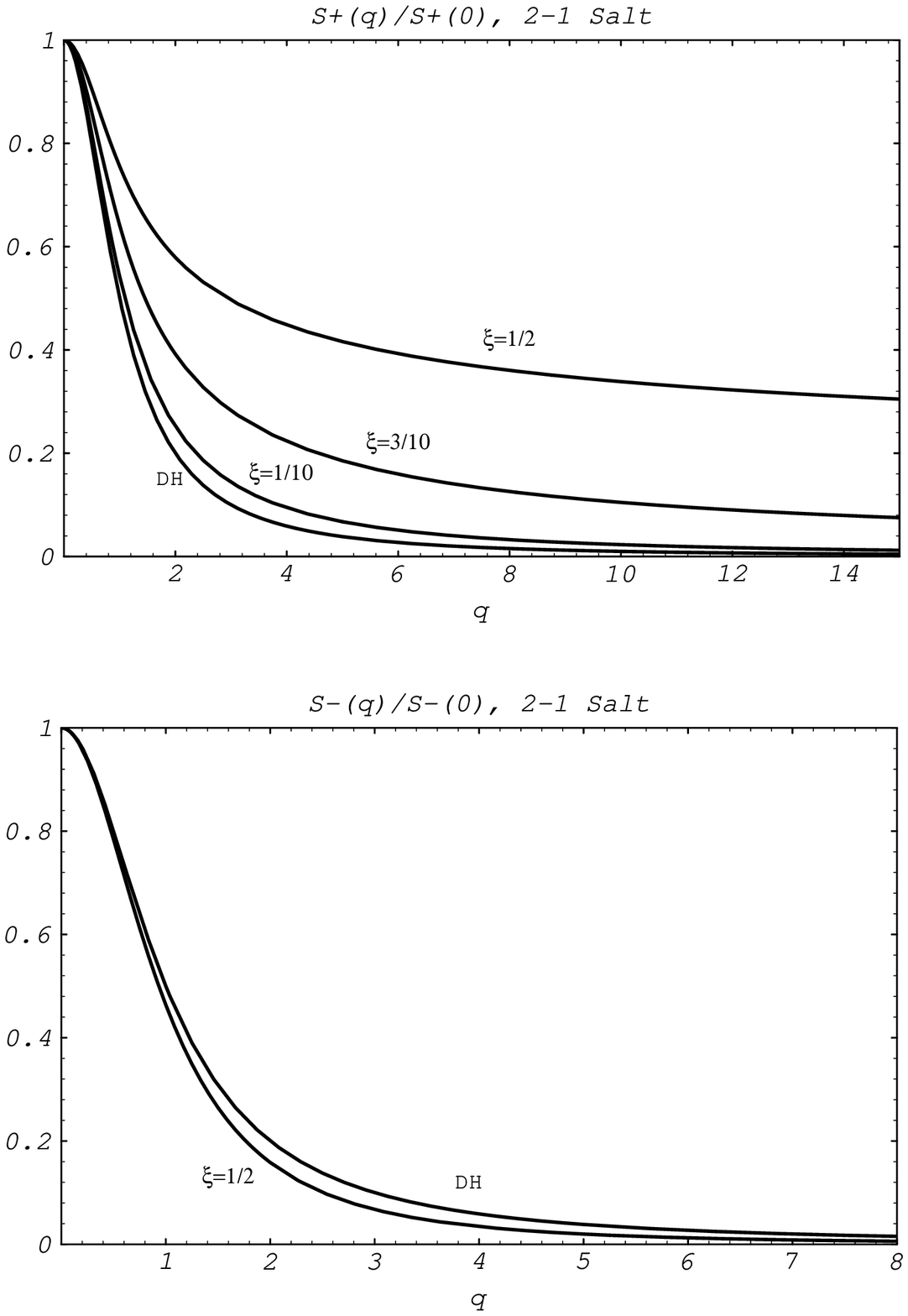}
\vspace*{-10mm}
\caption{The structure factors $S_\pm(q,\xi)/S_\pm(0,\xi)$ as functions of $q$
for
the 2-1 salt.
Also plotted is the Debye-H{\"u}ckel approximation.}
\efig
\par
To determine the large $q$ asymptotics of $S_\pm(q,\xi)$,   we use the fact
that if
\[ f(r) \sim \, r^\la \mbox{ as } r\ra 0, \]
then
\[ {\cal F}(f)(q) \sim  C(\la)\,  q^{-2-\la}\mbox{ as } q\ra\iy \]
with $C(\la)=-2^{2+\la}\sin(\pi\la/2)\,\Ga(1+\la/2)^2$.
{}From (\ref{smallr11}) and the above fact we deduce
for the 1--1 salt  that as $q\ra\iy$, for fixed $\xi<1$,
\bq S_\pm(q,\xi) \sim {C_\pm \ov q^{2\mp 2\xi}} \ \label{SAsy}\eq
where $C_\pm=\exp\left(\pm y_0(\xi)\right) C(\mp 2\xi)$.
At the critical value  $\xi=1$ the short distance asymptotics
of $e^{y_{11}}$ are no longer of the above form (see (\ref{rsmall11C})).  For
this case we find as $q\ra\iy$
\begin{eqnarray*}
 S_+(q,1)/S_+(0,1)&=& {4\ov 3}\,{ 1\ov \log q }-
 {8\log 2\ov 3}\,{ 1 \ov \log^2 q}+\mbox{O}({1\ov \log^3 q}),\\
 \noalign{\vskip 5pt}
 S_-(q,1)/S_-(0,1)&=&{2\log q\ov q^4} +{2(2\log 2 -1)\ov q^4} +\mbox{lower
order terms.}
\end{eqnarray*}
Similar large $q$ expansions hold for the 2--1 salt.

\par
In  critical light or neutron scattering from simple fluid or magnetic systems,
 the large $q$ behaviour of $S(q)$ defines the critical
exponent $\eta$ \cite{fisher,zinnJustin}.  For mean field theories of critical
scattering,
$\eta=0$.  We remark that  even
 though the PB equation is a mean field theory,
we have a nonzero ``$\eta$'' ($\eta=\pm 2\xi$, $\xi<1$, for the 1--1 salt)
 which is a reflection of the fact that the short distance potential is
the bare Coulomb potential.  As the critical value of the Manning parameter is
approached, the  Poisson-Boltzmann
theory predicts an enhanced scattering at large
wave numbers from the concentration fluctuations
of the counterions while at the same time predicting little
change in the scattering from the concentration fluctuations of the coions.
As is the case in critical scattering \cite{tm,sengers},
the measurement of
these   effects may well prove difficult.
\sp
\begin{center}{\bf Acknowledgements}\end{center}
This work was supported in part by
the National Science Foundation through grants DMS--9303413 and DMS--9424292.

\end{document}